\pdfoutput=1
\documentclass{article}
\usepackage{cite}
\usepackage[utf8]{inputenc}
\PassOptionsToPackage{hyphens}{url}
\usepackage{hyperref}
\usepackage{dirtytalk}
\usepackage[english]{babel}
\usepackage{authblk}
\newcommand{\email}[1]{
\href{mailto:#1}{\nolinkurl{#1}}
}

\usepackage[none]{hyphenat}

\title{The other side of the Coin: \\ Risks of the Libra Blockchain}

\author[1, 2]{Louis Abraham\footnote{\email{louis.abraham@yahoo.fr}}}
\author[3, 4]{Dominique Guégan\footnote{\email{dguegan@univ-paris1.fr}}}

\affil[1]{École polytechnique}
\affil[2]{ETH Zurich}
\affil[3]{University of Paris 1 Panth\'eon-Sorbonne}
\affil[4]{Ca' Foscari University of Venice}

\begin{document}
\sloppy

\maketitle

\begin{abstract}

Libra was presented as a cryptocurrency on June 18, 2019 by Facebook. On the same day, Facebook announced plans for Calibra, a subsidiary in charge of the development of an electronic wallet and financial services. In view of the primary risk of sovereignty posed by the creation of Libra, regulators and Central Banks quickly took very clear positions against the project and expressed a lot of questions focusing on regulation aspects and national sovereignty.
 
The purpose of this paper is to provide a holistic analysis of the project encompassing several aspects of its implementation and the issues it raises.
We address a set of questions that are part of the cryptocurrency environment and blockchain technology that support the Libra project.
We describe the governance of the project based on two levels, one for the Association and the other for the Libra Blockchain. We identify the main risks considering at the same time political, financial, economic, technological and  ethical risks.  We emphasize the difficulty to regulate such a project as it will depend on several countries whose legislations are very different. Finally, the future of this kind of projects is discussed through the emergence of Central Bank Digital Currencies.
\end{abstract}



\section{Introduction}

Libra was presented as a cryptocurrency on June 18, 2019 by Facebook. On the same day, Facebook announced plans for Calibra, a subsidiary in charge of the development of an electronic wallet and financial services.

\

To avoid a regulatory confusion around this project, the different banking and financial institutions, along with public actors, quickly reacted and created several task forces. They also asked Facebook to work with the appropriate regulatory authorities. Thus, as the host of the G7 and the G7 Finance Ministers meeting of Chantilly\footnote{\url{https://www.economie.gouv.fr/G7-finance-en}}, France put a task force in place.\footnote{\url{https://www.reuters.com/article/us-facebook-crypto-france/france-creates-g7-cryptocurrency-task-force-as-facebooks-libra-unsettles-governments-idUSKCN1TM0SO}\\
The goal of this task force is to create a regulatory frame encompassing crypto assets like Libra: drawing the first conclusions from the status report of the G7 working group to better assess the opportunities as but also the risks associated with this class of assets, particularly in terms of financial stability, transmission of monetary policy, consumer protection, money laundering and terrorist financing.}

Other institutions such as the Financial Stability Board (FSB), the European Banking Authority (EBA), Central Banks\footnote{The European Central Bank, headed by Benoit Coeuré, released a report on stable wedges at the end of 2019 (crypto currencies pegged to currencies or financial assets, see Section \ref{definitions}).}, the Financial Stability Board (FSB), the Bank for International Settlements (BIS) also reacted among others. As early as July, the American Senate Committee asked David Marcus (CEO of Calibra) to suspend his project, while Singapore is concerned \footnote{\url{www.mas.gov.sg/Singapore-Financial-Centre/Smart-Financial-Centre/Project-Ubin.aspx}} and India says it is not comfortable with the project \footnote{\url{https://economictimes.indiatimes.com/tech/internet/facebook-may-abort-libra-launch-in-india/articleshow/69867426.cms}}.

\

While the Libra Association sought an authorization in Switzerland from FINMA for Libra to be used as a means of payment, Bruno Le Maire, the French Minister of the Economy, announced on September 11, 2019, that \say{allowing the development of Libra on European soil} is out of the question. In his view, there are three issues at stake with the initiative of the social network: (1) the \say{monetary sovereignty of States}, (2) the fight against the illicit uses of money and finally (3) financial stability and risks for individuals and businesses.\footnote{\url{hhttps://www.businesstimes.com.sg/banking-finance/g-7-countries-seeking-common-stance-on-facebooks-libra}}.

In view of the primary risk of sovereignty posed by the creation of Libra, the Central Banks quickly took very clear positions. France and Germany have declared wanting to ban Libra\footnote{\url{https://www.reuters.com/article/us-facebook-cryptocurrency-france-german/france-and-germany-agree-to-block-facebooks-libra-idUSKCN1VY1XU}\\
\url{https://journalducoin.com/altcoins/reaction-libra-chine-accelere-projets-cryptomonnaie/}}. Head of the People's Bank of China's (PBoC) Research Bureau Wang Xin said that in the event of a massive adoption of Libra, this new currency would have the potential to profoundly alter monetary balances, financial stability and even the international economic system\footnote{\url{https://www.scmp.com/economy/china-economy/article/3017716/facebooks-libra-forcing-china-step-plans-its-own}}. In his statement, he indicated the possible adverse effects of this \say{made in USA} stablecoin. According to him, it is difficult to imagine that a digital currency backed by the dollar (the Libra) is totally equivalent and as powerful as a sovereign currency (the dollar itself), because \say{there is only one boss, namely the US dollar and the United States}. At the same time, we learned that China was going to issue a digital government currency (see Section \ref{definitions} on CBDCs). Other countries have also announced their intention to create their own digital currency: Iran plans to create a currency indexed on the Rial, Venezuela launched the Petro at the beginning of 2018 indexed on the country's reserve of oil, with 1 Petro having the price a barrel of oil, Russia talks about creating a currency indexed on oil\footnote{\url{https://www.ccn.com/russia-regulate-cryptop-with-a-keen-eye-on-oil-backed-digital-currency/}}, Marshall Islands are on the verge of issuing a digital asset after having adopted a law on February 26, 2018\footnote{\url{https://sov.foundation/}}, etc.

\

Nevertheless, at the same time, we can see that the Libra project is supported by powerful Silicon Valley groups.
Indeed, at the date of September 16, 2019, 28 entities accepted to support the launch of Libra, 25 of them investing 10 million dollars each (the 3 NGOs are exempted). Those entities include 6 payment services, 7 technology firms and marketplaces, 2 telecommunication firms, 4 Blockchain companies, 5 venture capital firms, 1 research center and 3 NGO. It is to be noted that 18 of those organizations are US-based and only 6 are European (including 2 UK firms). On October 4, 2019, Paypal announced their withdrawal from the project\footnote{\url{https://www.theverge.com/2019/10/4/20899310/facebook-libra-paypal-online-currency-payment-system-cryptocurrency}}. On October 11, 2019, Visa, Mastercard, Stripe, eBay and Mercado Pago followed\footnote{\url{https://www.theverge.com/2019/10/11/20910330/mastercard-stripe-ebay-facebook-libra-association-withdrawal-cryptocurrency}}.
Thus, every major US payment processor has exited the association and the only remaining payment provider is PayU, a Netherlands-based company. On October 14, 2019,  the internet company Booking left, just before the official launch of the Libra Association in Geneva with 21 founding members out of the original 28.\footnote{\url{https://www.cnbc.com/2019/10/14/facebook-forms-its-cryptocurrency-council-after-key-backers-drop-out.html}}\\

Over the few months since the original announcement, the Libra project has already undergone some changes. For example, an early version\footnote{\url{https://web.archive.org/web/20190618210315/https://libra.org/en-US/wp-content/uploads/sites/23/2019/06/TheLibraAssociation_en_US-1.pdf}} of a document describing the Libra Association \cite{libraassociation} described an \textit{Investment Token}, a security sold to early investors like the founding members of the Association that entitles their holder to voting rights and interests earned on the collateral. In the current version of the document, all references to this \textit{Investment Token} have disappeared, mentioning only the membership to the Association as a mechanism of governance.\\

This announced objective of the Libra project is financial inclusion, in order to make up for the lack of Central Banks in certain regions of the world. Libra proposes the creation of a currency allowing commercial transactions on different platforms in response to the demand of the unbanked or poorly banked. It seems difficult to deal only with the economic, or financial or political approach without asking the question of the technological environment that largely determines the project.

While there is an explosion of positions on the Libra project, mainly focused on aspects of regulation and national sovereignty, few documents analyze this phenomenon globally. We only see short statements, either positive, dubious, or anxiety inducing based on very specific aspects of the project. However, as we demonstrate in this paper, Libra cannot be restrictively viewed as a new system of money creation and transfers.

\

The purpose of our paper is to provide a holistic analysis of the project to encompass several aspects of its implementation and the issues it raises.
In this prospective work, we wish to address a set of questions that are part of the cryptocurrency environment and blockchain technology that support the Libra project. 
We are interested in identifying the main risks associated to this project in its preliminary phase. The questions we are interested in are: (i) What is Libra? Currency, crypto-asset, stable coin, settlement coin? (ii) Why are stable coins interesting and innovative: can they strengthen financial stability? (iii) What is new in the architecture of Libra? (iv) What are the risks of this project?
\\

The paper is organized as follows. In Section \ref{definitions} we recall some definitions and analyze the concept of stable coins and CBDC. We propose a first definition of Libra in the light of these notions. In Section \ref{technical} we analyze the technology behind the Libra project. Section \ref{regulatory-aspects} is devoted to regulatory matters. In Section \ref{risks} we analyze the different risks associated with this project and we conclude in Section \ref{conclusion}.

\section{Definitions/framework}
\label{definitions}

\subsection{Currency, monetary system, cryptocurrency}

In this section, we introduce the notions of money, cryptocurrency and stable coins. These different reminders allow us to analyze what could be the status of Libra currency.\\

A currency is defined by three functions: (i) unit of account for economic calculation or accounting, (ii) reserve of value and (iii) intermediary of the exchanges (ability to extinguish debts and obligations). A currency is characterized by the trust of its users in the persistence of its value and ability to serve as a means of exchange. It therefore has social, political, psychological, legal and economic dimensions. In times of unrest and loss of confidence, new currencies can appear.

The international monetary system is based on methods allowing the exchange of goods, debts and services between countries using different currencies. This system has often been subject to evolution. The crisis of 1929, considered responsible for the magnitude and duration of the depression, led to international cooperation, affirmed by the Bretton Woods agreements, introducing a fixed but adjustable exchange rate system, regulated by the International Monetary Fund (IMF). The abandonment of gold convertibility of the dollar in 1971 (Jamaica Agreement) led to a less regulated system of floating exchange rates. In 2008, the massive financial crisis that was observed has necessitated the organization of a more stable and regulated international monetary system. The Basel III agreements\footnote{\url{https://www.bis.org/publ/bcbs189.htm}} are one of the answers to make this system more stable. It is interesting to note that it was at this date that Satoshi Nakamoto (a pseudonym) \cite{nakamoto2008bitcoin} published the whitepaper that created Bitcoin and propagated the development of other crypto-assets based on a decentralized system without any trusted party. At that time, they believed that such an alternative financial system is the best way to counter the concentration of power and control in the too-big-to-fail financial institutions \cite{chen2019decentralized}.\\

Cryptocurrency (or crypto-asset) describes digital currencies that use cryptography to secure their payment networks and transactions in a decentralized way. Cryptography is a discipline dedicated to the protection of messages (confidentiality, authenticity, integrity) through the use of keys. This technique dates back to antiquity and was long considered an art and it was not until the 20th century that it became a science and the massive use of computers democratized its use.

This process ensures that the sender is the author of the message. Any digital currency based on a blockchain – a public ledger that records all transactions – which uses cryptography to encrypt recordings can be referred to as cryptocurrency. Bitcoin, Ethereum, Litecoin, Dash, Monero, and others fall under the category of cryptocurrencies. Other digital currencies allow secure, fast and low cost transactions and use blockchain technology, for example Ripple (XPR), but in this case, even if the protocol allows a decentralized management of the issue and associated transactions to this currency, there are only a few actors (validators), chosen by the company Ripple, who are authorized to operate on the network. Thus, Ripple, despite being based on blockchain technology, is not decentralized and is governed by a private blockchain. Ripple could be seen as a \say{simplified} representation of Libra in his governance but not in the technological approach as Libra does not create blocks to secure transactions (See Section \ref{technical}).\\

Currently, all crypto currencies are characterized by a high volatility relatively to their liquidity. This property that is interesting for the speculators makes their economic utility rather weak because economic actors are not interested in a currency that is too volatile. This is the reason why we have seen the introduction of the concept of stable coins in the cryptocurrencies universe since 2014.\\

In this paper, we designate by crypto-currency any digital currency that is secured by a blockchain-related protocol, public and totally decentralized type like Bitcoin, or private type or consortium like Ripple for example. In Appendix \ref{appendix-blockchains}, we give more information about these two concepts of blockchains, private or public, see also \cite{lin2017survey, guegan2018BMI151, guegan2018digital, cong2019blockchain}.

\subsection{Stable coins}

The principal idea for creating a stable currency is to use the Quantity Theory of Money (QTM) \cite{friedman1971theoretical} to determine when and how to print and destroy money. First, the notion of a currency's value relative to some ideal stable value must preexist. This value is most commonly specified by a Consumer Price Index (CPI), though most existing stable coins have chosen the USD as their measure of stable value. This choice of index of stable value is called the \textit{peg}. QTM is based on the relevant equation relating the price $P$ and the quantity of money $M$: $PM = \textrm{constant}$, which is derived from the classical equation of exchange.\\

Thus, to avoid the instability observed with other crypto currencies, Libra decides to create a reserve. In this project, the Libra reserve is composed of currencies equal to the entrance fees of the validators. The reserve is held in a basket of currencies, the repartition of which is known, and in principle the maintenance of the reserve is not more than the profit that can be achieved in the arbitrage \cite{librareserve}. Part of the profits will be used by the Association, and the rest will be paid to investors (Libra is a non-profit organization).\\

Any crypto-currency pegged to either a fiat currency or some kind of government-backed security (like a bond) counts as a stable coin. The idea is that this crypto-currency will be more \textit{stable} or less volatile. Asset-backed cryptocurrencies are not necessarily centralised, since there may be a network of decentralized vaults and commodity holders, rather than a controlling centralised body. The advantages of asset-backed cryptocurrencies are that coins are stabilized by assets that fluctuate outside of the cryptocurrency space reducing financial risk. Traditional crypto-currencies (Bitcoin, Ether, altcoins and others) are highly intercorrelated, so that cryptocurrency holders cannot escape global price falls without exiting the market of cryptocurrencies or taking refuge in asset-backed stablecoins. Furthermore, assuming that they are managed in good faith and have a mechanism for redeeming the assets backing them, asset-backed coins are unlikely to drop below the value of the underlying physical asset, thank to arbitrage.\\

We can distinguish several classes of stable coins even if the terminology is not fixed yet\footnote{A recent ECB paper \cite{bullmann2019search} proposes a possible classification. Another classification is proposed in \url{https://en.wikipedia.org/wiki/Stablecoin}, see also \cite{calcaterra2019stable, chohan2019stable}.}. 

\begin{itemize}

\item \textbf{Fiat-Collateralized stable coins}. The underlying model implies that the organization issuing a stablecoin holds a bank account containing the value of the issued tokens in fiat money. Thus, the organization can issue stable securities with a 1:1 ratio relative to its reserve in fiat currencies like the dollar or the euro, or other currencies. This model is fairly centralized knowing that it is the same entity who owns the assets represented. For example, the Tether currency is backed by the dollar (1: 1). Nevertheless, this one is quite controversial because it is impossible to verify if its reserve exists and if we are not in the presence of a Ponzi scheme.

The system related to Tether has evolved since they changed the backing to include loans to affiliate companies in March 2019. There are other crypto currencies backed by a single currency: USDCoin, Paxos, Gemini, Stably, ONRAMP, TrustUSD are all backed by the US dollar, Stasis is backed by the Euro, HKDK by the Hong Kong dollar, KRwb by the South Korean won \cite{calcaterra2019stable}. The advantage of these crypto currencies is that the price is stable and more resistant to the risks associated with the use of crypto currencies like Bitcoin, Ethereum, etc. The disadvantage is that it is a centralized system, the same actor controls the stablecoin and the reserve. For a stable coin to be attractive, it is important for the reserve to be publicly known and for the governance mechanism to be well defined. For example, Paxos Standard is a stablecoin whose US dollar reserve is held in various US banks. Despite the amount of the reserve in US dollars having been attested by independent auditors, the banks are not known.

Note that JPMorgan Chase \& Co.\ has officially become the first US bank to launch its own digital token representing a fiat currency (June 2019). The bank has announced the creation of JPM Coin, a blockchain-based technology that will facilitate the transfer of payments between institutional clients. The coin is backed by the US dollar with a 1:1 exchange ratio. For the bank, this JPM Coin is essentially a tool to help with the instantaneous transfer of payments between JPMorgan's clients, and the bank has indicated its plans to use JPM Coin for additional currencies as well in the future.

\item \textbf{Crypto-Collateralized stable coins}. In this model, the stablecoin is supported by other crypto-currencies. The question of volatility is still present but the system allows collateralization by absorbing price variations. Taking the example of the stablecoin Dai created by Markerdao and supported by the Ethereum, its principle is as follows: When Dais are generated, a smart contract is created and locks an amount in Ethers, a system that wants to be immutable. However, a major problem undermines this model: if the price of cryptocurrency used collapses, the certificates of issue could become 'sub-collateralized', which could then destabilize the price of the deposit. The stability of this cryptocurrency therefore depends on the health of another virtual currency. The advantages are that the system is decentralized, the liquidation can be faster and there is more transparency.

\item \textbf{Commodity-backed stable coins}. Stablecoins backed by commodities such as precious metals (gold, silver etc.) are much less likely to be inflated than fiat backed stablecoins. It is harder to mine gold or silver than it is to create money \say{out of thin air}. The main characteristics of these backed stablecoins are: (i) their value is fixed to one or more commodities and redeemable for such (more or less) on demand, (ii) there is a promise to pay, by unregulated individuals, agorist firms, or even regulated financial institutions. For example, Digix Gold is a currency backed by gold (1 gram of gold for 1 DGX), while PAXG is backed by one fine troy ounce ($1 \texttt{ oz t}\approx 31.1 \texttt{ g}$) of a 400 oz London Good Delivery gold bar, held in custody by the Paxos Trust Company. The questions around these stable crypto currencies are the following: (i) Does the amount of commodity used to back the stablecoin reflect the circulating supply of the stablecoin? (ii) Can holders of commodity-backed stablecoins redeem their stablecoins at the conversion rate to take possession of real assets? Another main point is that the cost of maintaining the stability of the stablecoin is the cost of storing and protecting the commodity backing. 

\item \textbf{Algorithmic stable coins or Seignorage Shares}. In this model, the issued stablecoin is supported by nothing other than its value over a given period. This model uses an algorithm that can be a smart contract. If the total demand for the stablecoin increases or decreases, then it will automatically change the number of chips in circulation to maintain a stable price. This mechanism can be considered similar to how Central Banks expand and contract monetary supply. They buy and sell fiscal debt to stabilize purchasing power. The advantages of this approach are that the system is decentralized and independent, adjustments are made on-line and no collateral is needed to mint coins. Finally, the value is controlled by supply and demand through algorithms that stabilize the price. The disadvantage is that the coin cannot be liquidated in case of problems. No such stable coins currently exists.

\end{itemize}

To conclude, the concept of stablecoin is at the experimental stage and those in circulation are still prototypes. However, they could in the future bring stability to the existing decentralized systems as means of decentralized payment. In a recent report\cite{coeure}, the G7 Working Group on Stable Coins led by Benoit Coeur\'e analyses the challenges of cross-border payments using so-called \textit{Global Stable Coins}. They make recommendations to help the financial regulators to \say{maintain a high level of international coordination and collaboration for cross-border policies that apply to stable coins}. \\

\subsection{The Libra stablecoin}

Libra is a stable cryptocurrency but unlike most of the others, it is not produced through a mining process. The transactions in Libra are done using a novel cryptographic state replication protocol. Thus, Libra, whose exchanges are based on a peer-to-peer system, is closer to a digital currency than to a crypto-currency. Libra will be created (1) in exchange for fiat currencies (a user buys Libra for fiat), (2) by investors (the more they encourage users to use Libra, the more they receive bonuses).\\

Libra can be considered as a stable coin because it will be backed by a basket of fiat currencies. In response to the German regulator Fabio De Masi\footnote{\url{https://finance.yahoo.com/news/facebook-libra-made-u-dollar-205718144.html}}, Facebook has communicated on September 23, 2019 the nature of the composition of Libra currency basket which will include: dollar representing 50\% of the basket, the euro for 18\%, the yen with 14\%, the pound with 11\% and finally a little Singapore dollars up to 7\%, no Chinese yuan. This last condition was indeed part of the requirements of the US Senate, especially Senator Mark Warner who was the spokesman for the US position: out of question that the Chinese national currency is part of the stablecoin currency panel. \\

It is interesting to note that the stable coin Libra is not emitted for the same reasons as the previous mentioned stable coins. Indeed, all the previous fiat-collateralized cited stable coins are issued by \say{crypto-native} firms whose roots are in the nascent sector. These stable coins have been created and used for investors to hedge against spikes in bitcoin's price, and as a means to trade crypto-currencies without using dollars, and also for arbitrage opportunities. However, Facebook is not a \textit{crypto-native} firm.\\

The choice of Libra to build a stable coin lies on the objective to ensure that the pegging mechanism works as intended, so that the Libra coin does not undergo any crazy crypto pumps \& dumps. Holders of Libra coin probably won't face any bigger risk than holding cash.\\

To attain this objective, the founders decided to create a reserve for achieving value preservation, backing each coin with a set of stable and liquid assets. The objective is to avoid speculative swings that are experienced by existing crypto-currencies. The white paper \cite{librareserve} indicates that \say{the reserve assets will be a collection of low-volatility assets, including cash and government securities from stable and reputable central banks}.\\

It is important to note also that backed stable coins are subject to the same volatility and risk associated with the backing asset. If the backed stablecoin is backed in a decentralized manner, then they are relatively safe from predation, but if there is a central vault, they may be robbed, of suffer loss of confidence. We will come back on this in Section \ref{risks}.\\

As the value of Libra is pegged to several currencies the question of the ratio is important: is the ratio proposed on September 23, 2019 definitive? How can Facebook or the Libra Association justify it? In any case, the amount of the currency used to back the stablecoin has to reflect the circulating supply of the stablecoin.\\


\subsection{Central Bank Digital Currency: CBDC}

CBDCs have been discussed for several years and the interest and statements around this topic were boosted by the announcement of the arrival of Libra \cite{kakushadze2018blockchain, coeure2018bitcoin, masciandaro2018cryptocurrencies}.\\

So far, the only public currency available to all citizens has been paper money and represents a significant share of the money supply in advanced economies. Nevertheless, there are some disadvantages to paper money: (i) it facilitates the growth of the illegal economy, with corresponding losses in terms of missing tax revenues, (ii) it undermines the effectiveness of monetary policy. But in return (i) the State acquires goods and services in exchange for banknotes, and (ii) the anonymity of the paper currency protects the privacy of the abuse of a state, be it democratic or dictatorial, to collect information using the payment system. We have just seen in the preceding paragraphs that crypto-currencies represent a new means of payment for individuals, produced and distributed according to a decentralized transfer system. Nevertheless, the use of crypto-currencies as a means of exchange is so far limited because of several characteristics of these currencies: their volatility, the slowness of their transactions, anonymity, the possibility of financing terrorism or money laundering, etc.\\

Beyond the resilience of traditional paper money, a natural question arises: is there a need to create a public digital currency (or currencies)? Such currencies (CBDC) could transform all aspects of the monetary system by serving as a no-cost-exchange system, as a secure store of value and as a stable unit of account for the benefit of consumers. Would it be interesting for Central Banks to issue digital assets? Some economists consider that it could increase GDP by reducing real interest rates, distortionary taxes and monetary transaction costs, and that it could improve the Central Banks ability to stabilize the business cycle, others describe efficiency gains in terms of payments, clearing and settlement due to the absence of anonymity \cite{barrdear2016macroeconomics, bech2017central}.\\

In countries where the Central Bank has little credibility, can a decentralized digital asset
help to stabilize the economy? Assuming that some digital assets become widely accepted, would it be possible to have bank accounts labeled in these digital assets? If so, how would the banks integrate these accounts in their risk model, e.g. to compute their Basel capital adequacy rate? Underlying these issues, there is the liability issue attached to the digital assets. Who is
responsible for any malfunction? Similarly, how to react to hacking? \\

The announcement of the creation of Libra has updated some of these issues. Like paper money, it would be a legal tender: the State would guarantee its role as a means of exchange (secure asset), but it would be distributed via centralized electronic networks. However, it would not have the anonymity property of the paper money which is distributed via
decentralized physical exchanges.\\

Obviously, a number of issues raised by the Libra project would also arise with a CBDC, even though the Central Bank would ensure the stability of the currency in that case.
For example, in such a system, all transactions are registered but the correspondence between user accounts and real-life identities is not a prerequisite. Thus, two conflicting points of concern arise, which are the privacy of users on one hand and the risk of money laundering on the other.\\

In this new environment, whatever protocol is considered, the issuer, the value of the currency and the means of distribution are to be taken into account. We analyze some of these points in the next sections in relation with the Libra project.

\section{Technical considerations on the Libra Blockchain}
\label{technical}

The \textit{Libra Blockchain} is defined as \cite{librablockchain}:

\begin{quote}
 a decentralized, programmable database designed to support a low-volatility cryptocurrency that will have the ability to serve as an efficient medium of exchange for billions of people around the world
\end{quote}

Libra is already implemented in an open-source prototype, \textit{Libra Core}, distributed under the Apache 2.0 License. Note that nowhere in this paper is the word \textit{blockchain} used without being part of the currency name.\\
  
  \subsection{A consortium blockchain}
  
  Unlike public blockchains, all users are not equal in Libra. A set of supernodes, called validators, are the only ones authorized to modify the database. The other nodes will be able to have a read-only access to that database. The validators are in charge of processing transactions and executing consensus algorithms to keep the database in a consistent state.
  Finally, the transactions are programmable (akin to the smart contracts introduced by \textit{Ethereum}) in a new programming language called \textit{Move}. In the first version, only predefined transactions are available but in the future it is said that users will be able to program their owns. However, no timeline is given regarding the implementation of this feature.
  \\

  The validators are the members of the Libra Association that governs the network and manages the reserve. Initially, the validators will be entities selected by Facebook and who paid \$10 million. It is said that \textit{membership eligibility will shift to become completely open and based only on the member's holdings of Libra}. This organisational system would be the administrative equivalent of a Proof-of-Stake, as it would probably be realized through a membership to the Swiss organization.
  However, in the paper dedicated to this issue \cite{libramoving}, no clear plan is described to achieve that goal. This undermines the description made by various media of Libra as a Proof-of-Stake system. \textbf{Currently, Libra does not include any Proof-of-Stake mechanism as it relies on a permission system.} The described governance mechanism is in fact more similar to a Proof-of-Authority (see Appendix \ref{appendix-protocols}).\\
  
  \subsection{Lifetime of a transaction}
  
  The typical interaction of a client wanting to execute a transaction is the following:
  \begin{enumerate}
   \item The client contacts a validator node to submit a transaction. This validator is named the \textit{leader} for that transaction.
   \item The validator proposes the transaction to the other validators.
   \item The network of validators executes the transaction and runs a consensus protocol.
   \item The consensus protocol outputs a signature that proves that the transaction was carried out.
  \end{enumerate}
  
  Hence, all transactions must first be proposed by a \textit{leader} who acts as an intermediary between the client and the network of validators. It is to be noted that there is no protection against censorship attack in the system as described. For example, a validator could refuse to treat a transaction and claim that they cannot execute it for the moment because of a global issue or just return an invalid answer.
  \\

  \subsection{Databases}
  
  We introduce the database of transactions, the queries and the user accounts and their relation with real-world entities.
  
  \subsubsection{The database of transactions}

  The data of the Libra Blockchain is constituted of a versioned ledger modified by transactions that produce outputs. Each ledger version is identified by a number equal to the number of transactions that have been executed. The ledger's state at time $i$ is noted $S_i$. Applying transaction $T_i$ to $S_{i-1}$ produces $S_i$ along with an output $O_i$. Thus, the entire database at time $n$ is constituted of the triplets $(S_i, T_i, O_i)$ for every $0 \le i \le n$.
  
  \
  
  Even the Libra paper states: \say{There is no concept of a \textit{block} of transactions in the ledger history.} How then can Libra be called a blockchain?\\
    
  \subsubsection{Queries}
 
  The whitepaper \cite{librablockchain} claims that clients will be able to issue queries to read data from the global database.
  When a transaction is executed, the database is modified and the new state is collectively signed by all or a part of the validators. This allows a client to verify the answer to a query without trusting the third party.
  
  Furthermore, the sparse Merkle tree data structure allows any client to perform authenticated read queries. Thus, those queries can be answered by any actor with a copy of the state, and this actor will be able to generate the cryptographic proof that the answer is correct. A similar mechanism allows untrusted third parties to provide feeds for events that can be cryptographically verified by any client.
  
  \
  
  Downloading the entire database should also be possible, and is said to allow the verification of past transactions. However, no mechanism is described to report or punish a dishonest validator through the Libra protocol.

  \subsubsection{User Accounts}
  \label{user-accounts}
  
  Libra uses an account system that authenticates users through public-key cryptography. An account can possess both data and code. For example, the data will contain the amount of Libra this account possesses.
  An account is not linked \textit{by the Libra protocol} to a real-world identity. However, the US Anti-Money Laundering (AML) and terrorism financing standards that must be met by Facebook will force the Libra exchanges to register the real-world identity of people when they change fiat currencies for Libra\footnote{\url{https://cointelegraph.com/news/libra-must-comply-with-anti-money-laundering-standards-us-treasury}}.
  For the existing crypto-currencies (like Bitcoin), most of the crypto-exchanges already record the banking information of their clients. However, since coin mining is possible in Bitcoin, anybody can in theory obtain new money that is not linked to any real-world identity. This problem does not exist in Libra as the only way to create new coins is to invest in the reserve of the Libra Association.
  
  \
  
  However, if a new user receives money from an existing user, it is not possible to know the real-world identity of the receiver unless the sender discloses it.
  Will Facebook require new accounts to be linked to a real-world identity? If they don't, the cost of any investigation for money laundering will skyrocket. If they do, it will give the members of the Libra Association a significant edge over the other nodes by allowing them to know the real-world identity of the accounts and aggregate the behavior of individuals.
  
  \
  
  Hence, it is probable that the mapping between real-world entities and Libra accounts will constitute another source of data, not stored by the Libra blockchain but by the Association members.
  
  Thus, Libra places itself in opposition to the very foundations of crypto-currencies, namely the cypherpunk and crypto anarchist movements aiming to overcome the centralization of banking services.

  \subsection{Programmable resources}
  
  Programs defined on the Libra blockchain are called \textit{modules} and are similar to \textit{Ethereum}'s smart contracts. They are represented in the Move bytecode \footnote{Bytecodes are a mid-level representation of a program: the highest level is the human-readable source code whereas the lowest level is the compiled program. A bytecode is typically executed by a virtual machine and for most existing languages, it is possible to easily produce human-readable source code from a bytecode representation.}. The Move language is not yet released and in the current version of Libra, a lower-level but human-readable language (called Move IR for intermediate representation) is used.\\
  
  Thus, it can be considered that any \textit{module} distributed through the Libra blockchain will be open-source. Because the Apache 2.0 license under which Libra Core is released is a permissive license, the Libra \textit{modules} can be licensed differently by their creators. What steps will Libra take to protect the intellectual property of their users?

  \subsubsection{Security}
  
  For the moment, no feature to update modules is implemented, which raises security concerns as it won't be possible to fix even known bugs before attackers find them.
  Attacks of smart contracts on the Ethereum platform are frequent \cite{atzei2017survey}. The most famous, The DAO attack, allowed hackers to steal 3.6 million Ether in a few hours, worth about \$50 million at the time and more than \$700 million now. Other errors of implementation can freeze smart contracts, making their assets unusable instead of stealing them. \\
  
  If accounts are linked to real-world identities, attacks will be less likely on the Libra Blockchain as legal sanctions would be applicable. However, this risk is not non-existent, with some countries not being bound by extradition treaties. Even worse, these bugs could be triggered inadvertently and unlike transferred money that can be sent back, burnt money is not normally recoverable.
  What guarantees will Libra offer to address smart contract bugs? For the moment, type-safety, reference-safety, and resource-safety are guaranteed by bytecode verification but it doesn't exclude. Are the Association or the validators likely to reverse some transactions? How will they decide to do so?\\
      
    \subsubsection{The \textit{gas} mechanism}
    
    To mitigate the risk of denial-of-service attacks, Libra uses the \textit{gas} mechanism introduced by Ethereum: every transaction is associated with a cost imputed to the client and expressed in gas. The fee associated with any transaction is proportional to the computing power used by the transaction and can be predicted (e.g. every operation has a fixed cost). However, the gas price is subject to variations over time: it rises when there are a lot of transactions to execute.
    The mechanism that decides the gas price is not detailed in the paper. A consequence of the gas system is that validators are supposed to treat transactions with higher gas price first and are allowed to dismiss some of the others.\\
    
    Since the gas price is not publicly determined by the protocol or by a trusted actor, validators are free to \say{drop transactions with low prices when the system is congested} \cite{librablockchain}. This means that the level of the fees can change arbitrarily and dynamically and this process is not clear for the users. The user has no guarantee that his transaction will be done, depending on the fees they propose. This raises concerns over competition mechanisms that will come into play. A rule capping the gas price in normal load situations should therefore be put in place, to prevent consumer isolation and ensure the liquidity of Libra.
    
    \subsubsection{Storage capabilities}
    
    Beside computation power, another limited resource is memory. The paper mentions the possibility of a similar gas system to limit storage. Even in the absence of such mechanism, a denial-of-service attack based on short-term memory would be costly: the memory usage of any program is always limited (up to a constant) by the number of operations it makes. Instead, the real cost of the memory is for long-term storage.
    
    \
    
    Thus, a rent-based system along with a policy to define data expiration times would be more adapted to limit the storage used by transactions and is briefly described. In this mechanism, a fee in Libra would be charged and proportional to the size of the data and the time during which it has to be accessible. However, in that system, the validator nodes are said to be able to recache the contents of a transaction after the expiration time. Thus, the validators still keep a copy of any data.\\
    
    This memory management system raises further concerns regarding data protection laws. For example, no company targeting European customers should use a Libra smart contract to store personal data because of the \textit{right to erasure} (see Section \ref{data-protection}).\\

  \subsection{The Byzantine consensus protocol}
  
  At the core of Libra is the concept of Byzantine fault tolerance, which allows components of a distributed system to fail unpredictably without disrupting the operation of the whole system.\\
  
  The Byzantine consensus protocol of Libra, LibraBFT, is based on HotStuff and similar to a lot of modern protocols. The collective decisions are all authenticated using \textit{Quorum Certificates}, as in most Byzantine agreement algorithms. Such QCs are collective signatures that ensure that two thirds of the validators agreed to a transaction. The particularity of the QCs used by Libra is that they preserve the identity of the validators who signed. Thus, the cryptographic internals of the signature mechanism are problematic to determine the liability of the validators. If an unlawful transaction was validated, the responsibility would probably be shared by all the validators as there is no way to know the identity of the validators who accepted it.\\
  
  \subsection{Authorized resellers}
  
  The service of conversion between fiat currencies and Libra is planned to be ensured by \textit{authorized resellers}.\\
  
  As with any secured system, the weakest spot is its interactions with other systems. Authorized resellers will be able to indirectly control the minting and burning of coins by exchanging fiat and Libra coins and asking the validators to adapt the volume of coins to the demand. How are the communications between the resellers and the Libra association secured and authenticated? What measures are put in place to ensure the trustworthiness of the resellers? How is the risk that some resellers go bankrupt managed?
  
  \subsection{Latency and resiliency}

  The blockchain is said to support up to 1,000 transactions per second with 10-second intervals between transactions \cite{librablockchain}: is it sufficient to be competitive against other systems? It seems that the volume of exchanges will be much lower than what is offered by the other classical payments like Visa, Mastercard or Paypal who withdrew from the project in October 2019.\\
  
  We also learn in the same paper that validators will be able to store up to 4 billions accounts and that they are free to discard historical data to be able to process new transactions. What is the insurance for a client that their transactions have not been discarded? Nothing is said about consumer protection or traceability.

\section{Regulatory aspects}
\label{regulatory-aspects}

In the previous Section, we have addressed some questions arising from the protocol itself, mainly about the protection of the customers (account, transaction, privacy, security, etc.).
In this Section, we address some other regulatory aspects concerning the Libra project. 
First, we describe the dual governance system of the Association and of the protocol. We then ask questions about the applicability of an external jurisdiction. We also approach the regulation of Libra as a platform of payment. Finally, we analyse the applicability of data protection directives like the GDPR to Libra.

  \subsection{Governance}

  A centralized organization is naturally given a hierarchy that is liable and must deal with any situation that does not achieve the chosen objective. In the decentralized world, this structure does not always exist and rules have to be put in place to reduce bad behaviors and negative outcomes. Thus, specifying the governance system becomes crucial to control the actions of the ecosystem actors.\\
  
  It is important to distinguish between \textbf{two different levels of governance} that are relevant for Libra: the traditional governance of the Libra Association based in Switzerland, and the algorithmic governance of the Libra Blockchain yet to be launched.
  
  \subsubsection{Governance of the Libra Association}
  
  The governance of the Libra Association is similar to a circle of investors who have paid the amount required to belong to the Association. This Association manages the code (development, update, search, extension of the system, ...).
  
  
  \
  
  Concerning the governance of the Association, in an interview with senators on July 16, 2019, David Marcus (CEO of Calibra) declared: \say{you'll have this council of a hundred or more members that will make decisions and will elect a board that will be between five to nineteen members and that board will then of course elect a managing director and the Association will have a staff to perform the governance function that it is supposed to look after}\footnote{\url{https://youtu.be/xUQpmEjgFAU?t=4294}}. Nevertheless, nothing is specified for the moment on the way the decisions will be taken: What is the voting system of the Association? How do new members enter the Association? On other side, who can leave the network and how?\\
  
  Concerning possibly malicious members, what rules are to be applied? Can they be banned? What are the consequences for the management of the Reserve? Are the mechanisms put in place by Libra sufficiently clear, supposing they exist?\\

  \subsubsection{Governance of the Libra Blockchain}
  
  
  It seems that Libra wants to implement some sort of algorithmic governance: \say{[The Libra Association] can create Move modules that delegate the authority to mint and burn coins to an operational arm that interacts with authorized resellers.} \cite{librablockchain}. Thus, using Move, they will be able to use \say{flexible governance mechanisms such as this will allow the Council to assert its authority and take over its delegated authority through a vote}. It is not clear how the members of the Association are concerned in the decision of the management of the Reserve which is a key point for this project \cite{librablockchain}.\\
  
  Distributed governance is based on a \textit{flat-hierarchy} philosophy and on the idea that token holders will dedicate sufficient time to participate and vote according to the interests of the company. To strengthen the investor involvement, many projects have thus developed incentives which can be summarized as a \textit{monetization of attention} methodology, by granting rewards (like tokens) to acquire the attention of a network of investors. What powers will be given to validators? Their role need to be clearly identified and defined in the Libra project.
  
  A conflict between validators could happen. How does the protocol react to avoid collusion, misappropriation of funds or non-compliance with rules to make transactions in a continuous and non-discriminatory way?
  
  In fact, we observe that in the first version of Libra, the validation process will be controlled by the founding members of the Libra Association. Then, the Libra project aims to move toward a decentralized governance architecture. However, no clear plan or technical protocol is proposed to achieve this objective.\\

  This new governance process that combines traditional and algorithmic governance has not been totally analyzed and opens interesting questions. 
  Indeed, governance of decision-making bodies plays a crucial role in the stability design. The stability mechanism will be automated in a day-by-day functioning on the network, but monetary and fiscal policies, like the management of the Reserve, cannot be fully automated.\\


  \subsection{External jurisdiction}
  
  The biggest obstacle to the legislation of Blockchain technologies is that they are traditionally decentralized by essence.
  However, Libra, as a stable coin, aims to represent a new milestone in the adoption of Blockchain technologies. The pegging mechanism they introduce is based on a centralized governance by the Libra Association. As such, regulations could be applied to this entity.
  This Association is based in Switzerland. However, all members are from foreign countries, with a majority of American members.\\
  
  From the European point of view, it is already a matter of defining the object to be regulated: is it an electronic money as per the Electronic Money Directive (EMD2), a financial instrument as per the Market in Financial Instruments Directive (MiFID), a fund as per the Payment Service Directive (PSD2)\footnote{ \say{Directive 2004/39/EC}. Official Journal of the European Union. 2004. Retrieved 20 March 2008.}, or an object of a new nature that does not constitute a regular activity under European law?\\
  
  In some aspects, the technological nature of this project can get in the way of the regulators. For example, in the event of an inquiry about a wrongful transaction validated through the Libra protocol, all validators (currently the members of the Association) would be involved because the anonymous signature mechanism protects the identity of the validators who accepted that transaction.

  \subsection{Libra as a payment platform} 
  
    \subsubsection{Authorization}
    
    In order to be able to play the role of payment platforms, the Libra payment platform must seek the authorization of the financial authorities. They filed such a request to the FIMNA\footnote{Financial Market Supervisory Authority} in Switzerland on September 11, 2019.
    Simultaneously, the President of the United States, Donald Trump, wrote on his Twitter account: \say{If Facebook and other companies want to become a bank, they must seek a new Banking Charter and become subject to all Banking Regulations, just like other Banks, both National and International.}\footnote{\url{https://twitter.com/realdonaldtrump/status/1149472284702208000}}.\\
    
    How is it possible for Libra to meet the demand of regulators when the rules in terms of regulation are very different from one country to another? What will be the trust of countries in that currency and the transactions done with it? May there be an embargo on the Libra from some countries?\\
    
    It is important that all countries adopt the same policy regarding Libra: if one country authorizes Libra as an asset, the conversion to all currencies will be possible by propagation.

    \subsubsection{Consumer protection}
    \label{consumer-protection}
    
    At the infrastructure level, what would happen if the system was unavailable (whatever the cause of the unavailability), thus interrupting all transactions? The economy of one or more countries would be blocked. To whom the customers of this system could turn to case of injury? Several courts of justice would be concerned. We already see many disparities in the current legislation regarding the status of blockchain \cite{blemus2019initial, de2018blockchain}. How can these issues be understood in the framework of Libra?\\
    
    Will the principles of scalability developed in the Move code be able to avoid traffic jams: it is proposed to increase the level of fees in case of high demand. How to be sure that will be enough? Would customers have insurance in terms of maximum time for the transfer of their requests?\\
    
    What would happen in case of cyber-attack (or just in case of rumors of this type)? One can easily foresee a panic phenomena on the part of holders of Libra who will try to sell their assets, potentially leading to a risk of \textit{bank run}.\\
    
    \subsubsection{Know Your Customer and Anti-Money Laundering}

    The fight against corruption, especially in the banking system, is not a new subject. The Foreign Corrupt Practices Act (FCPA), although published in 1977 in a very turbulent environment marked by the Watergate case, remains more relevant than ever. Its implementation remains a joint priority of the Department of Justice (DOJ) and the Securities and Exchange Commission (SEC) in the United States. The UK also published very strong regulation, the UK Bribery Act in July 2011, which is recognized as one of the most demanding anti-corruption laws.\\
    
    In 1989 the G7 established the Financial Action Task Force (FATF) and decided that \say{When a company acts for a customer, it must know the beneficial owner of the operation, that is to say the natural person(s) who own or control effectively a client and/or the natural person on whose behalf a transaction is done} \cite{ferrari2019compliance}. \\
    
    Recently, authorities in the US and abroad have increased their focus on modernizing and enforcing anti-money laundering and terrorism financing (AML) regulations. As part of these efforts, the US's Financial Crimes Enforcement Network (FinCEN) proposed Know Your Customer (KYC) requirements in 2014, which were finalized in 2016. FinCEN's KYC requirements were proposed as part of a broader regulation setting out the core elements of a customer due diligence program. Taken together, these elements are intended to help financial institutions to avoid illicit transactions by improving their view of their clients' identities and business relationships.\footnote{\url{https://corpgov.law.harvard.edu/2016/02/07/fincen-know-your-customer-requirements/}}\\
    
    Concomitantly, Europe adopted the Directive 2014/95 / EU of the European Parliament and of the Council on October 22, 2014 amending Directive 2013/34 / EU concerning \say{the publication of non-financial information and information on the diversity of their activities by certain large companies and groups}. Following this directive, for example in France, the Sapin II law passed in 2016 and imposes that companies with an international dimension comply with certain national and international laws and regulations linked to the fight against corruption in relation with extraterritoriality rules in particular\footnote{The Sapin II Law (2016). 2016-1691 of 9 December 2016.}.\\
    
    The concepts of extra-territoriality are associated with the Foreign Corrupt Practices Act (FCPA) to combat bribery of public officials abroad. It covers all acts of corruption by companies or persons, US or non-US, that are either located in the US, or simply listed on US stock exchanges, or who participate in a way or another to a regulated financial market in the United States. It is implemented by the Office of Foreign Assets Control.\\
    
    Countries have also tightened their legislation, and supervisory institutions are responsible for these issues, the TracFin, and CNIL in France, in England this concerns the FSA, La Bafin in Germany, in Singapore the MAS, etc.
    
    Thus, each financial institution has to indulge in the Know-Your-Customer (KYC) process with customers to comply with regulations, such as Anti-Money Laundering (AML) and Countering the Financing of Terrorism (CFT). Each Line of Business within a financial institution performs their own customer checks. The customer typically provides KYC documents each time he requires services within an institution. How is Libra going to solve to protect the customer and provide the information at the institutions responsible of the supervision relative to KYC, AML and CFT? As part of the implementation of a regulation, it will be necessary to establish a system verifying that the rules governing the KYC process and the AML regulations are respected. What entity, independent of Libra, could play this role?


  \subsection{Compliance with Data Protection Directives}
  \label{data-protection}

  We first remind the reader that \textbf{Libra possesses data at at least two very different levels}: the blockchain transaction data that are public and the user account data (see section \ref{user-accounts}) that will be owned by companies. The former is limited and difficult to regulate while the regulatory efforts should be concentrated on the latter. Some third-party services might also emerge around Libra, we also put them in the second category. In this part, we mainly address the regulation of the first category, the blockchain transaction data, as it is the most original and difficult to comprehend.
  
  \
  
  In Europe, the General Data Protection Regulation (GDPR)\footnote{ \say{Official Journal L 119/2016}. eur-lex.europa.eu. Archived from the original on 22 November 2018. Retrieved 26 May 2018.} that came into application on May 2018 has the objective to reach a balance between data protection and legitimate use of data. The GDPR aims among other things to take into account the significant technological developments on the past 25 years, keeping their goal of the protection of the individual rights. The database technology that enables the decentralization of data storage and processing through blockchain seems to be difficult to interpret in the frame of the GDPR. If those challenges seem difficult with public blockchain, in the case of a private permissioned network, the requirements could be easier to interpret and implement. They have to be identified and integrated in the process. We list some of the obligations created by the GDPR and try to understand if Libra has taken them into account.
  
  \
 
  Some issues that Libra needs to address to be compliant with the GDPR rules concern: (i) the identification of the identity and obligations of data controllers and data processors, (ii) the identification of the processings that are necessary for the performance of the service (article 6.1 of the GDPR), (iii) the exercise of some data subject rights: it could be difficult to rectify or remove personal data from the network (right to erasure). Thus, the main point for regulators is to identify a data controller having lawful grounds to collect the information as defined in the GDPR, must be transparent about how it intends to use the data and must ensure that he does nothing unlawful with them. \\
  
  This data controller also has the responsibility to safeguard the personal data they collect \say{whether against hacks or against accidental loss, destruction and damage}\footnote{\say{EUR-Lex – 32016R0679 – EN – EUR-Lex}. eur-lex.europa.eu. Archived from the original on 17 March 2018. Retrieved 21 March 2018.}. The personal data \say{shall be collected for specified, explicit and legitimate purposes}, and only used for the purposes that have been stated, and also data \say{must be kept in a form which permits identification for data subjects for no longer than necessary} for the stated purposed.\\
  
  In the case of Libra, where validators are part of the governance, rules need to be precisely defined to decide who is able to see the data and who is going to play the role of data controller or of joint controllers. In any case each network participant should agree to some terms and conditions relative to GDPR directives before being granted access to the network.\\
  
  An important fact is that, because of the decentralized nature of Libra, the data will be shared with nodes all over the world. Hence, the GDPR rules regarding data transfers outside the EU apply to any data present on the Libra Blockchain.\\
  
  The regulation of other countries can somehow get in the way. For example, how is it possible to reconcile the GDPR and the CLOUD (Clarifying Lawful Overseas Use of Data) Act? The former asks for respect of the data ownership where as the latter clarifies the rules allowing the US authorities to requisition data stored outside their territory by U.S.-based technology companies. We know that this American regulation poses difficulties in terms of respect for privacy, especially for Europeans. What solution does Libra propose to answer at the same time the requests of the GDPR and the CLOUD Act?\\
  

\section{Risks and Libra's project}
\label{risks}

What is the real objective of the Libra project? The Libra project is at the opposite of a decentralized system even if it is presented as using a peer-to-peer system. Decentralized systems facilitate innovation and competition to produce newer, better and cheaper financial services: this aspect is not clear with this project. Indeed, it is said that the project can move to this decentralisation, but how to be sure of this evolution and how will it work?\\

Any project based on a blockchain-related innovative system creates new risks we need to analyze. This project is challenging because it combines many risks for which answers and strategies must be developed. We can distinguish several kind of risks (without being exhaustive): financial, economic, political, technological, operational, ethic, privacy, etc.\ even if the barrier between all these risks is not strict. As soon as the project was brought up in June 2019, some of these risks have been identified. It is interesting to verify whether Libra addresses them in the documents that are available. \\

  \subsection{Political risks}
  
  The political risk is probably the first one that the national institutions and politics mentioned as soon as the project has been announced. Some concerns are: 
   \begin{itemize}
    \item \textbf{Monetary sovereignty}. We can consider this risk from three different approaches. (i) What is the impact of Libra on existing fiat currencies? (ii) Does the Libra project influence or activate the decision of creation of CBDCs by some countries? (iii) Can the creation of a supra-national currency impact the global financial system?
    
    \textbf{(i)} What would be the impact on the economy of a country if the use of Libra were to reach a significant percentage of the monetary exchanges, knowing that Libra is backed 50\% on the US dollar? Could we imagine sanctions coming from the USA on commercial exchanges in Libra with some countries against which the American government decided an embargo? Conversely, would a country whose economy is partially dependent on the Libra be able to support economic sanctions against the US?
    
    
    Moreover, the sovereign status of any national currency seems protected because it is used to pay taxes. Can one imagine that this evolves in the future, and that it will be possible to pay taxes in Libra?
    
    

    \textbf{(ii)} The proposal of creation of Central Bank Digital Currencies, in response to the rise of cryptocurrencies, is not new as we know that Central Banks are in the process of creating their own cryptocurrencies-stablecoins pegged to their own fiat currencies. Thus, in the future, we will see the emergence of several national CBDCs. What will be their relationship with fiat currencies? A new exchange market will emerge creating some new instability. What will be the success of these stable coins? For sovereignty reasons, States must have a perfect knowledge of the technology they use and completely control their protocol.
      
    \textbf{(iii)} Is Libra really a danger for the global financial system?  Other systems have already been put in place in some countries to unify payments and act as an interface between transactions: it is the case for instance in India with WhatsApp. However, WhatsApp Pay depends on the Indian government's Unified Payments Interface to handle the transactions. Other companies develop similar services like Humaniq or Kiva, but they have not as much money or technological resources as Facebook. On another side, can Libra disrupt Visa or Mastercard? It is not probably the goal of this project. Their geographical influence area will not correspond to the regions in which Facebook is interested to develop this new ecosystem. In any case, it will be important to analyse and understand how these different systems can competitively exist. 
    For instance, what will happen to the stable coins currently in use like Tether and Dai with the arrival of Libra and CBDCs?
    
    \item \textbf{Adoption by emerging countries}. Some countries, particularly emerging countries, are ready to support this project, as access to financial services is still limited for some groups of people. As most emerging countries probably won't have the technical means to create their own digital currency in the near future, should Libra bear this new avatar of the \say{white man's burden}? Indeed, one can consider that the new Libra ecosystem could address some of these issues, serving as platform for innovative financial services, like seamlessly sending money. Should people use Libra instead of their national currencies in developing countries, what would be the impact on the national economies? Would they plummet in favor of international exchanges?

    \item \textbf{Partial adoption}. A risk would be the emergence of two blocks of countries, one rejecting the project and its use, and the other block accepting this currency. This situation would be an element of global friction and create more financial instability than stability. One of the major consequences would be the difficulty of setting up trade flows smoothly. From a macro-economic point of view, what would be the sustainability of a world monetary system, with fiat transactions on one side and transactions using an oligopolistic type of currency like Libra on the other? Thus, we argue that the ability of a single country to authorize or ban Libra at its level is at most an illusion, and that the practicability of a unified regulatory framework will be a determinant factor in the decision process.


   \end{itemize}
   
  \subsection{Financial risks}
  Even if the Libra's project is not finalized and has to be postponed due to the recent demands of regulators, institutions and banking system, we can discuss several questions related to the position of Libra in the landscape of the financial industry.\\ 
  
  First of all, it is important to recall the specificities of a bank: they are institutions that fall under a regulatory umbrella covering institutions that offer insured deposits and their holding companies. The regulatory apparatus that restricts the actions of banks was built largely because banks offer deposits, which bring both value and fragility \cite{diamond2000theory}. To mitigate the risk, banks are subject to regulatory costs. For instance, banks abide by many regulations that force them to take extra steps to make sure that their customers are not using their services for money laundering. Banks are also required to hold minimum amounts of capital to satisfy existing regulations. We currently observe that FinTech firms do not have to follow the same regulations \cite{buchak2018fintech}. It is interesting to understand how Libra will be positioned in relation to the obligations related to the creation of money, deposits, etc.
  
  \begin{itemize}
   \item \textbf{Libra as a bank}. Can Libra have the status of Central Bank? If Libra finalizes their project, a stable digital currency will be created along with a reserve created by the staking of the investors. However, the management of monetary policies must remain the private turf of Central Banks for reasons of national sovereignty and financial stability. Thus, the regulators need to clarify how this project can emerge.
   Since Libra challenges the financial status quo, it will also be of importance for Central Banks to respond to the innovation brought by this project.
   
   \item \textbf{Financial stability}. Does the system proposed by Libra increase financial stability or in return create more uncertainty for the world population? Libra wants to be a fast, easy, cheap and secure payment system to ensure liquidity. The current protocol seems to solve this question even if we note in section \ref{technical} a certain number of questions regarding access, security, etc. How can we be sure that this system will not evolve to become more restrictive, creating a gap between the expectations of the user and the final objective of the project?
   
   \item \textbf{Convertibility}. What is the expected exchanges rate of Libra? The importance of convertibility of Libra to other currencies need to be explained in detail for the users. The political and regulatory choices regarding Libra will not be probably the same for countries all around the world, and could be changed at any period of time. Thus, a currency risk can emerge and will impact for instance the possibility to buy groceries or other things with other currencies than Libra.
   
   \item \textbf{Bank run} could happen in case of cyber-risks, frauds, panic, etc. If a lot of people decide to change their Libra (congestion risk) to fiat or CBDC currencies, it could cause a systemic risk, with a cascade of failures for the investors associated to the project. In terms of management risk, how has this kind of crisis been taken into account by the Libra project? Is the existence of the Reserve sufficient to avoid this crisis? What would happen in the event that a major actor leaves the network, like this happened with Paypal, Visa, Mastercard and Stripe?
   
   The safety of a bank in a traditional system is controlled by the regulator following very specific steps, and it is based on the computation of a capital requirement encompassing the risks the bank faces. What equivalent system will Libra propose to avoid such failure? What guarantees does this project offer? Libra promises to keep a fiat reserve matching the Libra supply. However, no plan for accountability is provided. Will Libra, as a banking activity, be subject to capital requirements defined by regulators?
  
  \end{itemize}
   
  \subsection{Economic risks}
  
  In relation to the financial stability, the economy of countries can be impacted by the creation of a new currency. Not repeating the definition or the status of Libra as a currency, we indicate some new situations that can appear with the emergence of this project.
  
   \begin{itemize}
   \item \textbf{Antitrust concerns}. Does the Libra system promote the creation of an oligopolistic market?\footnote{Oligopoly is a market structure with a small number of firms, none of which can keep the others from having significant influence.}.
   
   As early as August 2019, Europe questioned the potentially anti-competitive behavior of Libra\footnote{\url{https://www.bloomberg.com/news/articles/2019-08-20/facebook-s-libra-currency-gets-european-union-antitrust-scrutiny}}.
   The status of Facebook as advertiser, which is already the motive of antitrust investigations including one in Europe \footnote{\url{https://www.bloomberg.com/news/articles/2019-07-02/facebook-is-latest-to-come-under-eu-s-antitrust-scrutiny}} and one conducted by 47 US state attorneys general\footnote{\url{https://ag.ny.gov/press-release/2019/attorney-general-james-gives-update-facebook-antitrust-investigation}}, strengthens those concerns. Indeed, a widely missing feature in online advertising is the ability to track whether an advertisement has led to a purchase. Such an advertisement would be sold for a higher price and could grant the advertiser a commission on the sale. With Libra, it will be child's play for Facebook to track the user's expenses and thus widen the gap with its competitors.
   
   
   From another point of view, what will competition between systems based on stable crypto-currencies look like? What will be the equilibrium between a system based on both fiat and stable coins and a system using only stable coins? Is a macro economic model able to explain in a robust way this equilibrium and what are the constraints to attain it? The literature on the subject matter suggests that the balance might be difficult to find \cite{cales2018theoretical, sanchez2019truthful}.

    
   \item \textbf{Taxation}. How will tax authorities come up to Libra? Backing Libra into a currency basket does not protect it from exchange rate fluctuations. As soon as capital gains emerge, taxes will apply. In the US, the current IRS law treats digital currency as property, so every single transaction or purchase made with Libra will be a taxable event. In France, in November 2018, the taxation of cryptocurrencies was aligned with that of securities. The highest values are taxed at 30\% according to the flat tax mechanism. Hence, Libra holders will have to take into account the tax regulations of each country. When transferring Libra or goods, there will be also the need to trace and report annually all payments for the calculation of these taxes. What does Libra plan to ease the work of the fiscal administrations? How clearly will taxpayers be informed of these obligations?
   
   \end{itemize}
   
  \subsection{Technological risks}
  
  The creation of a new protocol opens the doors to new questions and risks. We have already listed some of them when we analysed this protocol in Section \ref{technical}. Since the protocol is not yet in production, all risks are not identified, but we discuss some of them.

  \begin{itemize}
  
  \item \textbf{Unplanned program behaviors}. We know that problems can arise, such as operational failures, code vulnerabilities, malicious programs, etc. The ability to solve them fast will make the system even more robust. 
  
  \item \textbf{Operational risks}. They concern cyberattacks, fraud, hacking, etc.\ and are linked to the security of the protocol. Are the procedures put in place to limit these risks clearly explained?

  \item \textbf{Data collection}. These questions are already addressed in Section \ref{regulatory-aspects}, we come back to them because they are fundamental: Where and how are customer data stored and who has access to them? What are the authentication methods used? How has Libra planned to implement the laws on anti money laundering? Where are the user account data (see section \ref{user-accounts}) stored, in order to check if KYC is respected?
  
  \item \textbf{Personal data}. In practical terms, when considering the use of obfuscation, encryption and aggregation techniques to process personal data, one must consider two risks: (i) the risk that misuse reduces the strength of the cryptographic algorithms (we suppose that the target security levels are respected) (ii) the risk that encrypted data can be linked back to an individual by examining patterns and contexts of usage, possibly by aggregation with other information sources. Furthermore, the use of AML/KYC data for other purposes must be closely monitored and prevented.

  \item \textbf{Updates of the protocol}. How can the protocol updates be assessed? How frequent will the protocol be updated? This point is important for regulation, hence the need — as for AI algorithms — to have an accurate, clear, detailed documentation to know this information in due time.\footnote{Remember that the protocol that we analyze today (October 2019) is a prototype. It will probably evolve before deployment.
  }
  
  \item \textbf{Systemic risk}. What happens upon failure of the Libra protocol? A platform outage could be caused by various phenomenon (bug, internet outage, censorship, power outage, etc.). It would deprive all users of their means of payment, purchase, transfers, communication. What solutions does Libra provide to address these extreme failure cases? However, in the area of systemic risks, decentralized systems can be considered to be more reliable than centralized ones.
  
  \item \textbf{Interoperability}. If a user wants to use another wallet than Calibra, will it be possible or easy to export their keys and Calibra data to another platform? In his hearing before the US Senate, David Marcus announced that Calibra would be interoperable with WhatsApp. What about other platforms? We note that the GDPR introduced a \say{right to portability}.

  \end{itemize}

  \subsection{Ethic and regulation}
  We cannot analyze the Libra phenomenon without talking about ethics. We have covered this topic several times throughout the paper. We come back and specify some points that we think are important.
   \begin{itemize}
   
    \item \textbf{Bank status}. If Libra has a status close to that of a bank, it is essential for it to be subject to the same rules as other banks in order to avoid bankruptcy. This raises the question of the risk measure to be used which will encompass all the risks associated with this new environment. All regulators adopting a common policy for calculating this financial risk will be crucial. Can the Basel III directives be applied? What adjustments must be made to integrate stable coins into risk calculations?
    
     \item \textbf{Data property}. The question of data ownership is at the heart of the project. How can the Libra system be made compliant with the rules of the GDPR? How can two systems as complex and antithetical as the CLOUD Act and the GDPR co-exist? 
     
     \item \textbf{Consumer protection}. What guarantees does the consumer have in case of a stolen account, transactions that are not made, or piracy of the platform? Can we imagine that the proposed protocol would go against the law (e.g. no refund in case of theft)?
     
    

    \label{moral-hazard}
    \item \textbf{Moral hazard} is a legitimate concern, as most of the assets controlled by the Libra Association will be deposited by the users. Indeed, the double nature of Libra as a reserve of value and a mean of payment is involved as the real risk bearers will be the users who entrusted Libra with the management of their assets and use it as a payment platform. However, this risk can be minimized in the presence of a regulatory supervision of the project.
    
    Moreover, the global misinformation regarding this project (including its widely undisputed status of cryptocurrency) creates an information asymmetry between the Libra project and its users who are attracted by the cryptocurrency fad. Finally, most users probably won't be in the know about the collusion between Facebook's advertiser interests and the Libra project.
    Therefore, more effort should be dedicated to raise the consumers' awareness about the inner workings of online advertising.
   \end{itemize}

\section{Conclusion} 
\label{conclusion}



Regardless of the purpose of the Libra project, all the discussions and official statements clearly indicate that, for future similar projects, the constraints that will be imposed on confidentiality, money laundering, consumer protection and financial stability will have to be taken into account from the outset in the drafting of the protocol and the governance of the project. For Libra to launch, a good compromise is yet to be found between the novelty necessary to attract customers and the legal constraints that are not negotiable. The announcement of Libra as a crypto-currency serves both purposes: it rides the wave of enthusiasm towards crypto-currencies and scares potential regulators with technical matters. On the contrary, we rather tend to think of Libra as a digital currency.\\

A major lesson that can be drawn from the reaction of the different actors and our analysis is that regulating Libra will be a hard task since it has to be carried out on a global level. First, the double governance of the system must be taken into account: there is an organisational scale through the Libra Association and a network scale through the consensus protocol. Hence, the technical aspect must not become a barrier for non-expert regulators who already possess the competences to regulate the Libra Association. However, the technology will also have to be taken into account in a second time. This will represent a significant endeavor for the regulators, given the extent of the confusion brought along by the Libra project, but even what seems to be details can prove to be crucial to assess the risks. For example, the anonymity of the nodes who validated a transaction is a property of the cryptography used in the Quorum Certificates, and will make establishing
the liability of a single actor impossible in the event of fraudulent transactions.\\

Most worryingly, the vagueness of the non-technical aspects of the project is without precedent. Up until now, Facebook seems to have focused their efforts on the technical aspect of their project and overlooked the absence of clear rules for the governance of the network. Notwithstanding the political, economic and ethical aspects, the biggest technological risks are not internal dysfunctions of the Libra Blockchain but lie in its interactions with the outside world that are not specified: consumer protection, KYC/AML, authorized resellers, etc. Furthermore, there is a clear mismatch between Facebook's objective through Libra and the consumer's interests as most people ignore the mechanisms of online advertising.  \\

Nevertheless, an open question is the desirability of a \say{synthetic hegemonic currency}\footnote{\url{https://www.bankofengland.co.uk/-/media/boe/files/speech/2019/the-growing-challenges-for-monetary-policy-speech-by-mark-carney.pdf}} replacing the dollar in international payments, as described by Mark Carney, the outgoing Governor of the Bank of England. Indeed, the currently dollar-dominated international monetary system raises concerns regarding the dependence of any international exchange on the dollar, the ensuing liquidity shortage, the pressure on falling interest rates on government bonds and the uncertainty of the Central Banks which are dependent on the US Federal Reserve. What stable currency could replace the dollar? Probably not the Libra that is backed by national currencies and that will not be able to provide liquidity in times of crisis. It will be up to the national banks to solve this last problem. In addition, an international currency also has the property of being backed by a fiscal authority able to collect taxes, which will never be the case of Libra. The issuance of CBDCs might prove a solution to this problem.

\section*{Acknowledgements}

We would like to thank Louis-Samuel Pilcer, Sayuli Drouard and Cyril Grunspan for valuable discussions and comments.


\pagebreak

\bibliography{ref}{}
\bibliographystyle{plain}

\pagebreak

\appendix

\section{Public and Private blockchains}
\label{appendix-blockchains}

Cryptocurrencies like Bitcoin are electronic money based on a peer-to-peer or decentralized system using cryptography to validate transactions and also to create money. The protocol is a decentralized fiduciary mechanism to avoid the use of a trusted third party. The principle of decentralization implies that everyone can participate in the drafting of the code (it must nevertheless obtain an entrance fee). The blockchain associated to cryptocurrencies like Bitcoin is the sequence of all transactions. The use of cryptography ensures the security of transactions and their routing throughout the world.\\

The blockchain technology through which the transactions are executed is a technology of storage and transmission of information, transparent, secure, and functioning without a central control body. By extension, a blockchain is a database that contains the history of all the exchanges made between its users since its creation. This database is secure and distributed: it is shared by its different users, without intermediaries, which allows everyone to check the validity of the chain. Such a blockchain is a public blockchain.\\

Transactions between network users are grouped in blocks. Each block is validated by the nodes of the network called the miners, according to techniques that depend on the type of blockchain. In the Bitcoin blockchain, this technique is called \say{Proof-of-Work}, and consists in solving algorithmic problems using an important computing power with consensus principle by all nodes present on the network. Once the block is validated, it is time-stamped and added to the blockchain. The transaction is then visible to the receiver as well as the entire network.\\

We can distinguish three main types of blockchain system, public blockchain, consortium and private blockchains. Public blockchains are decentralized and available to anyone with an internet connection, no one has control over the network and they are secure in that the data can't be changed once validated on the blockchain: an example is the Bitcoin blockchain.  Private blockchains are typically used in enterprise solutions to solve business cases and underpin corporate software solutions. They are said to be permissioned:  participants need to obtain an invitation or permission to join the network, an example is the R3 Corda blockchain. The consortium blockchain is a \say{semi-private} system with a controlled user group, but works across different organizations. We classify Libra as a consortium blockchain.

\section{Different protocols}
\label{appendix-protocols}

The transfer and creation of crypto-currencies lie on different protocols. These protocols allow to make the transactions secure and to verify that they can be done. Several kinds of protocols exist: Proof-of-Work (PoW) protocols have been used for instance for Bitcoin while Proof-of-Stake (PoS), the Delegate Proof-of-Stake (dPoS) and Proof-of-Authority (PoA) protocols are alternative protocols to PoW.\\

\textbf{Proof-of-Work} is used by Bitcoin and a lot of other crypto-currencies to create crypto-currencies and transfer them. PoW is a specification that defines an expensive computation, called mining, that needs to be performed in order to create a new group of trusted transactions (the so-called block) on a distributed ledger called blockchain. Mining serves to verify the legitimacy of a transaction, avoiding the so-called double-spending problem, and to regulate money creation by rewarding miners for performing the previous task. To get the rewards, miners solve a mathematical puzzle known as Proof-of-Work problem. In most PoW protocols, this \say{mathematical puzzle} is an operation of inverse hashing: it determines a number (nonce) such that the cryptographic hash algorithm of block data is less than a given threshold. This threshold, called difficulty, is what determine the competitive nature of mining: the more computing power is added to the network, the higher this parameter increases, increasing also the average number of calculations needed to create a new block. This method also increases the cost of block creation, pushing miners to improve the efficiency of their mining systems to maintain a positive economic balance. \\
 
\textbf{Proof-of-Stake} is a different way to validate transactions and achieve distributed consensus. It can also be called Proof-of-Participation (PoP). This algorithm has the same purpose as the Proof-of-Work, but the process to reach the goal is quite different. In Proof-of-Stake, the creator of a new block is chosen depending on their wealth, also defined as stake. There is no block reward so the miners, now called forgers, only earn from transaction fees. Thanks to PoS, validators do not have to use a lot of computing power because the only factors that influence their chances are the total number of their own coins and current complexity of the network. The security of the system is based on the fact that if a validator creates an invalid block, their security deposit will be deleted as well as their privilege to be part of the consensus network.\\

\textbf{Delegated Proof-of-Stake} (DPoS) protocols are a subcategory of the basic Proof-of-Stake protocols, typically used in consortium blockchain systems. In these protocols, blocks are minted by a predetermined set of users of the system called delegates, who are rewarded for their duty and are punished for malicious behaviour (such as participation in double-spending attacks). In DPoS algorithms, delegates participate in two separate processes: (i) building a block of transactions, (ii) verifying the validity of the generated block by digitally signing it. While a block is created by a single user, it typically needs to be signed by more than one delegate to be considered valid. The list of users eligible for signing blocks changes periodically using predefined rules. \\

In \textbf{Proof-of-Authority}-based (PoA) networks, transactions and blocks are validated by approved accounts, known as validators.[1] PoA is comparable to a \say{Proof of Stake model that leverages identity as the form of stake rather than actually staking tokens}.\footnote{\url{https://blockonomi.com/proof-of-authority/}} Validators run software allowing them to put transactions in blocks. The process is automated and does not require validators to be constantly monitoring their computers. However, it does require maintaining the computer (the authority node) uncompromised. With PoA, individuals earn the right to become validators, so there is an incentive to retain the position that they have gained. By attaching a reputation to identity, validators are incentivized to uphold the transaction process, as they do not wish to have their identities attached to a negative reputation. PoA is suited for both private and public networks. The term was coined by Gavin Wood, co-founder of Ethereum and Parity Technologies.[2]\\

In summary, PoW is based on the computing power of the participants, PoS on the wealth of the participants, and PoA on the reputation of the participants.


\end{document}